\documentclass[aps,prd,showpacs,twocolumn,nofootinbib]{revtex4}
\usepackage{graphicx}
\usepackage{dcolumn}
\usepackage{bm}

\begin{document}

\def\be{\begin{equation}}
\def\ee{\end{equation}}
\def\bd{\begin{displaymath}}
\def\ed{\end{displaymath}}
\def\ba{\begin{eqnarray}}
\def\ea{\end{eqnarray}}
\def\lr{\leftrightarrow}
\def\nn{$n\bar n$ }
\def\qq{$q\bar q$ }
\def\cc{$c\bar c$ }
\def\ccbar{$c\bar c$ }
\def\x{\bf x}
\def\B{\rm B}
\def\D{\rm D}
\def\E{\rm E}
\def\F{\rm F}
\def\G{\rm G}
\def\H{\rm H}
\def\I{\rm I}
\def\J{\rm J}
\def\K{\rm K}
\def\L{\rm L}
\def\M{\rm M}
\def\P{\rm P}
\def\S{\rm S}
\def\T{\rm T}
\def\V{\rm V}
\title{\ Bottomonium Spectrum with Screened Potential}
\author{
Bai-Qing Li$^{a,b}$ and Kuang-Ta Chao$^a$}

\affiliation{$^a$Department of Physics and State Key Laboratory of
Nuclear Physics and Technology, Peking University, Beijing 100871,
China; \\$^b$Department of Physics, Huzhou Teachers College, Huzhou
313000, China}

\date{\today}

\begin{abstract}
As a sister work of Ref.~\cite{Li:2009zu}, we incorporate the
color-screening effect due to light quark pair creation into the
heavy quark-antiquark potential, and investigate the effects of
screened potential on the spectrum of bottomonium. We calculate the
masses, electromagnetic decays, and E1 transitions of bottomonium
states. We find that the fine splittings between $\chi_{bJ}$
(J=0,1,2) states are in good agreement with experimental data, and
the E1 transition rates of $\Upsilon(2S)\to\gamma\chi_{bJ}(1P)$ and
$\Upsilon(3S)\to\gamma\chi_{bJ}(2P)$ (J=0,1,2) all agree with data
within experimental errors. In particular, the mass of
$\Upsilon(6S)$ is lowered down to match that of the
$\Upsilon(11020)$, which is smaller than the predictions of the
linear potential models by more than 100 MeV. Comparison between
charmonium and bottomonium in some related problems is also
discussed.

\begin{flushleft}
\end{flushleft}
\end{abstract}

\maketitle
\section{Introduction}
Potential models have been successful in describing the spectra
below the open-flavor thresholds for both charmonia and bottomonia.
However, it is well-known that these quenched potential models,
which incorporates a Coulomb term at short distances and a linear
confining potential at large
distances~\cite{Eichten:1978tg,Eichten:1979ms,Godfrey:1985xj}, will
overestimate the masses of heavy quarkonia above the open-flavor
thresholds. Some distinct examples are the $\chi_{c2}{'}/Z(3930)$ in
the charmonium system and the $\Upsilon(6S)$ in the bottomonium
system, of which the observed masses are about 50 and 90$\,MeV$
respectively lower than that predicted by the
 typical relativized potential model of Godfrey and Isgur~\cite{Godfrey:1985xj},
 and even $120\,MeV$ lower for $\Upsilon(6S)$
 than the prediction of the Cornell model~\cite{Eichten:1978tg,Eichten:1979ms}.

This is probably because,  the linear potential, which is expected
to be dominant at large distances, is screened or softened by the
vacuum polarization effect of the dynamical light quark
pairs~\cite{Laermann:1986pu}. This  screening or string breaking
effect has been demonstrated indirectly by the investigation of the
mixing of static heavy quark-antiquark ($Q\bar Q$) string with a
static heavy-light meson-antimeson ($Q\bar q$-$\bar Q q$) system in
the $n_f=2$ lattice QCD calculations~\cite{Bali:2005fu}, and has
also been implied  recently by the calculations within some
holographic QCD models~\cite{Armoni08:Screening}.

However, since the simulations of lattice QCD still have large
uncertainties and difficulties in handling higher excited states, it
should be useful to improve the potential model itself to
incorporate the screening effect and compare the model predictions
with experimental data, as a phenomenological way to investigate the
screening effects on heavy quarkonium spectrum.

Such screened potential
models~\cite{chao1989,Chao:1992et,Ding:1993uy,Ding:1995he} were
proposed many years ago in the study of heavy quarkonium and heavy
flavor mesons, as well as light hadrons\cite{ZhangZY}. The main
feature of these screened potential models in the spectrum is that
the masses of the higher excited states are lowered.

In recent years the screened potential models have again been used
to investigate the heavy quarkonium spectrum and leptonic decay
widths ~\cite{Segovia-etal:ScreenedPotential}. In
Ref.\cite{Li:2009zu} we have reinvestigated the charmonium spectrum
within the screened potential model suggested by Chao and
Liu~\cite{chao1989} and assigned some newly discovered
charmonium-like resonances as conventional higher charmonium states.

On the experimental side, aside from abundant  resonances discovered
recently in the charmonium region, progress in the bottomium region
has also been made. The $\Upsilon(1D)$ was observed by CLEO
collaboration~\cite{Bonvicini:2004yj} in 2004 and the $\eta_b$ was
observed by BaBar collaboration~\cite{:2008vj} in 2008. One may
expect more bottomium states will be observed in the future by
BaBar, Belle and CLEO. So it is important to reinvestigate the
bottomonium system within the screened potential model.

In this paper, as a sister work of \cite{Li:2009zu}, we calculate
the mass spectrum and electromagnetic decay and transition rates of
bottomonium especially the higher bottomonium using a
non-relativistic Schr\"{o}dinger equation with the Coulomb potential
plus a screened linear potential, which is nearly the same as that
in~\cite{Li:2009zu}. The model predictions  are similar to that of
\cite{Li:2009zu} for charmonium. The mass of $\Upsilon(6S)$ is
lowered to be consistent with its experimental value.

In the following, we first introduce the screened potential model in
Sec.II, and then study the mass spectrum and decay and transition
processes of bottomonia in Sec.~III.  In Sec.~IV we will discuss
some features of our result for the bottomonium states. A summary
will be given in Sec.~V.

\section{The screened potential model}
\label{Themodel} As a minimal model describing the bottomonium
spectrum we use a non-relativistic potential model with the
screening effect being considered as in~\cite{Li:2009zu}. We use a
potential as
\begin{equation}
 V_{scr}(r) = V_{V}(r) + V_{S}(r),
\end{equation}
\begin{equation}
 V_{V}(r) = -\frac{4}{3}\frac{\alpha_C}{r},
\end{equation}
\begin{equation}
 V_{S}(r) = \lambda (\frac{1-e^{-\mu r}}{\mu})+\,C.
\end{equation}
Here $\mu$ is the screening factor which makes the long-range scalar
potential of $V_{scr}(r)$ become flat when $r \gg \frac{1}{\mu}$ and
still linearly rising when $r \ll \frac{1}{\mu}$, and $\lambda$ is
the linear potential slope (the string tension), which is taken to
be the same as for charmonium\cite{Li:2009zu}. $V_{V}(r)$ represents
the vector-like one-gluon exchange potential, $\alpha_C$ is the
coefficient of the Coulomb potential. C is a constant related to the
normalization of energy levels of the $Q \bar Q$ system.


The spin-dependent interactions include three parts as follows. The
spin-spin contact hyperfine interaction is
\begin{equation}
\label{Hss}
 H_{SS} =\frac{32\pi\alpha_C}{9 m_b^2}\, \tilde \delta_{\sigma}(r)\, \vec
{S}_b \cdot \vec {S}_{\bar b}\, ,
\end{equation}
where $\tilde \delta_{\sigma}(r)$ is usually taken to be $\delta
(\vec r\,)$ in nonrelativistic potential models, but here we take
$\tilde \delta_{\sigma}(r) = (\sigma/\sqrt{\pi})^3\, e^{-\sigma^2
r^2}$ as in Ref.\cite{Barnes2005} since it is an artifact of an
O$(v_b^2/c^2)$ expansion of the T-matrix~\cite{Barnes1982} in a
range comparable to $1/m_b$.

The spin-orbit term and the tensor term take the common forms
\begin{equation}
\label{Hls}
 H_{LS} = \frac{1}{2m_{b}^{2}r} (3V_{V}^{'}(r)- V_{S}^{'}(r)) \vec
 {L} \cdot \vec {S},
\end{equation}
 and
\begin{equation}
\label{Ht}
 H_{T} = \frac{1}{12m_{b}^{2}}(\frac{1}{r}V_{V}^{'}(r)-V_{V}^{''}(r)) T.
\end{equation}
 These spin-dependent interactions are dealt with perturbatively. They are
diagonal in a $|J,L,S>$ basis with the matrix elements
\begin{equation}
 < \vec {S}_b \cdot \vec {S}_{\bar b} > =
\frac{1}{2} S^{2}-\frac{3}{4},
\end{equation}
\begin{equation}
< \vec {L} \cdot \vec {S} > = [J(J+1) - L(L+1) - S(S+1)]/2
\end{equation}
and the tensor operator T has nonvanishing diagonal matrix elements
only between $L>0$ spin-triplet states, which are
\begin{eqnarray}
<^3L_J |T| ^3L_J> =\left\{\begin{array}{ll}
   -\frac{L}{6(2L + 3)}\,,&J = L + 1\\\rule{0cm}{0.7cm}\frac{1}{6}\,,&J = L\\\rule{0cm}{0.7cm}-\frac{(L + 1)}{6(2L - 1)}\,,&J = L-1.
\end{array}\right.
\end{eqnarray}
 For the model parameters, we take
\begin{eqnarray}
 \label{para}
 \alpha_C=0.37,\alpha_{S}=0.18, \nonumber\\
\mu=0.056\,GeV,C=0.677\,GeV,\nonumber\\
m_b=4.4\,GeV,\sigma=3.3\,GeV,\lambda=0.21GeV^2
\end{eqnarray}
where $\alpha_C\approx \alpha_{s}(m_b v_b)$ and $\alpha_{S}\approx
\alpha_{s}(2m_b)$ are essentially the strong coupling constants but
at different scales. The former is for large distances and used to
determine the spectrum while the latter is for short-distances and
used for QCD radiative corrections in bottomonium decays (see below
in next section).

Here $\mu$ is the characteristic  scale for color screening, and
$1/{\mu}$ is about $3.5\,fm$, implying that at distances larger than
$1/{\mu}$ the static color source in the $b\bar b$ system gradually
becomes neutralized by the produced light quark pair, and string
breaking emerges. Note that here $\mu$ is smaller than that of
charmonium in~\cite{Li:2009zu}, where $\mu=0.0979\,GeV$
corresponding to $2\,fm$. In Sec.IV we will discuss the reason for
the difference in $\mu$ between $b\bar b$ and $c\bar c$ systems.

With these values of the parameters for the potential, we can
calculate the spectrum of bottomonium. The results are shown in
Table~\ref{spectrum_table}. For comparison, we also list the
experimental values~\cite{PDG08} and those predicted by the linear
potential model~\cite{Godfrey:1985xj} in Table~\ref{spectrum_table}.

\section{Some decay processes}

\subsection{Leptonic decays} The electronic decay width of the vector meson
is given by the Van Royen-Weisskopf formula~\cite{VanRoyen:1967nq}
with QCD radiative corrections taken into
account~\cite{Barbieri:1979be}.
\begin{equation}
\label{Swave:ee}
\Gamma_{ee}(nS)=\frac{4\alpha^{2}e_{b}^{2}}{M_{nS}^{2}}
|R_{nS}(0)|^{2} (1-\frac{16}{3}\frac{\alpha_{S}}{\pi}),
\end{equation}
\begin{equation}
\label{Dwave:ee} \Gamma_{ee}(nD) =
\frac{25\alpha^{2}e_{b}^{2}}{2M_{nD}^{2}m_{b}^{4}}
|R_{nD}^{''}(0)|^{2} (1-\frac{16}{3}\frac{\alpha_{S}}{\pi}),
\end{equation}
where $M_{nS}(M_{nD})$ is the mass for $nS(nD)$, $e_{b}=\frac{1}{3}$
is the $b$ quark charge, $\alpha$ is the fine structure constant,
$R_{nS}(0)$ is the radial S wave-function
 at the origin, and $R_{nD}^{''}(0)$ is the second derivative of the
radial D wave-function at the origin.

With the chosen parameters~(\ref{para}), we get the results that are
listed in table~\ref{tab:ee}. We also list other two groups'
results~\cite{Anisovich:2005jp,Pandya:2008vf} for comparison. We can
see that our results are consistent with the experimental data.

\subsection{Two-photon decays} In the nonrelativistic limit, the
two-photon decay widths of the ${}^1S_{0}$, ${}^3P_{0}$, and
${}^3P_{2}$ states can be written as~\cite{kmrr}
\begin{eqnarray}
\label{2gs0}
 \Gamma^{NR}({}^1S_0\to
 \gamma\gamma)&=&\frac{3\alpha^2e_b^4|R_{nS}(0)|^2}{m_b^2}\,,\\
\label{2gp0}
\Gamma^{NR}({}^3P_0\to\gamma\gamma)&=&\frac{27\alpha^2e_b^4|R'_{nP}(0)|^2}{m_b^4}\,,\\
\label{2gp2}
\Gamma^{NR}({}^3P_2\to\gamma\gamma)&=&\frac{36\alpha^2e_b^4|R'_{nP}(0)|^2}{5m_b^4}.
\end{eqnarray}

The first-order QCD radiative corrections to the two-photon decay
rates can be accounted for as~\cite{kmrr}
\begin{eqnarray}
\Gamma({}^1\!S_0\to\gamma\gamma)&=&\Gamma^{NR}({}^1\!S_0\to
\gamma\gamma)[1+\frac{\alpha_S}{\pi}(\frac{\pi^2}3-\frac{20}3)]\,,\rule{0.5cm}{0cm}\\
\Gamma({}^3\!P_0\to \gamma\gamma)&=& \Gamma^{\rm NR}({}^3\!P_0\to
\gamma\gamma)[1+\frac{\alpha_S}{\pi}
(\frac{\pi^2}3-\frac{28}9)]\,,\rule{0.5cm}{0cm}\\
\Gamma({}^3\!P_2\to \gamma\gamma)&=& \Gamma^{\rm NR}({}^3\!P_2\to
\gamma\gamma)[1 -\frac{16}3\frac{\alpha_S}{\pi} ]\,.
\end{eqnarray}

We can see that $\Gamma({}^1\!S_0\to\gamma\gamma)\propto
|R_{nS}(0)|^2$, which are sensitive to the details of the potential
near the origin. So we take
\begin{equation}
\Gamma({}^1\!S_0\to\gamma\gamma)\longrightarrow
\frac{\Gamma({}^1\!S_0\to\gamma\gamma)}{\Gamma_{ee}(nS)}\Gamma^{expt}_{ee}(nS)
\end{equation}
to eliminate this uncertainty.

In the nonrelativistic limit, we can also replace $m_{b}$ by M/2,
where M is the mass of the corresponding  bottomonium state. The
results are listed in Table~\ref {twophoton}. Predictions of some
other models (see
Refs.~\cite{Godfrey:1985xj,Ebert:2003mu,Munz:1996hb,Bhaduri:1981pn,Resag:1994ki,Gupta:1996ak,Schuler:1997yw})
 are listed for comparison. We can see our results are a
bit larger than most models but are consistent with
Refs.~\cite{Gupta:1996ak,Schuler:1997yw}.

\subsection{E1 transitions}

For the E1 transitions within the bottomonium system, we use the
formula of Ref.~\cite{Kwong:1988ae}:
\begin{eqnarray}
&&\Gamma_{\rm E1}( {\rm n}\, {}^{2{\S}+1}{\rm L}_{\J} \to {\rm n}'\,
{}^{2{\S}'+1}{\rm L}'_{{\J}'} + \gamma)\nonumber\\
&&=\frac{4}{3}\, C_{fi}\, \delta_{{\S}{\S}'} \, e_b^2 \, \alpha \,
|\,\langle f | \, r \, |\, i \rangle\, |^2 \, {\rm E}_{\gamma}^3 \,
\end{eqnarray}
 where E$_{\gamma}$ is the emitted photon energy.

The spatial matrix element
\begin{equation}
<f|r|i>=\int_{0}^{\infty}R_{f}(r)R_{i}(r)r^{3}dr\,,
\end{equation}
involves the initial and final state radial wave functions, and the
angular matrix element $C_{fi}$ is
\begin{equation}
C_{fi}=\hbox{max}({\L},\; {\L}')\; (2{\J}' + 1) \left\{ { {{\L}'
\atop {\J}} {{\J}' \atop {\L}} {{\S} \atop 1}  } \right\}^2 .
\end{equation}

Our results are listed in Table~\ref{E1rad}. The widths calculated
by the zeroth-order wave functions are marked by $SNR_0$ and those
by the first-order relativistically corrected wave functions are
marked by $SNR_1$.

For the first-order relativistic corrections to the wave functions,
we include the spin-dependent part of
(\ref{Hss}),(\ref{Hls}),(\ref{Ht}) and the spin-independent part
as~\cite{Miller:1982ca}
\begin{eqnarray}
H_{SI}&=&-\frac{\vec{P}^4}{4m_b^3}+\frac{1}{4m_b^2}\bigtriangledown ^2V_V(r)\nonumber\\
&&-\frac{1}{2m_b^2}\left\{\left\{
\vec{P}_1\cdot V_V(r) \Im \cdot \vec{P}_2 \right\}\right\}\nonumber\\
    &&+\frac{1}{2m_b^2}\left\{\left\{\vec{P}_1\cdot \vec{r}\frac{V_V^{'}(r)}{r}\vec{r}\cdot \vec{P}_2
    \right\}\right\},
\end{eqnarray}
where $\vec{P}_1$ and $\vec{P}_2$ are the momenta of $b$ and $\bar
b$ quarks in the rest frame of bottomonium, respectively, which
satisfy $\vec{P}_1=-\vec{P}_2=\vec{P}$, $\Im$ is the unit
second-order tensor, and $\{\{\quad\}\}$ is the Gromes's notation
\begin{equation}
\{\{\vec{A}\cdot \Re\cdot
\vec{B}\}\}=\frac{1}{4}(\vec{A}\vec{B}:\Re+\vec{A}\cdot \Re
\vec{B}+\vec{B}\cdot \Re \vec{A}+\Re:\vec{A}\vec{B}),
\end{equation}
where $\Re$ is any second-order tensor.

Note that we do not include the contributions from the scalar
potential in $H_{SI}$ since it is still unclear how to deal with the
spin-independent corrections arising from the scalar potential
theoretically.

We also list in Table~\ref{E1rad} the results of
Ref.\cite{Kwong:1988ae} which uses a potential obtained from the
inverse-scattering method for comparison.

Both the $SNR_0$ and Ref.\cite{Kwong:1988ae} results of E1
transitions are larger than most of the experimental values, but we
see that in $SNR_1$ the predicted widths get decreased and fit the
experimental values quite well as long as the first-order
relativistic corrections to the wave functions are taken into
account.

\section{Discussions}
\label{Discussion}
\subsection{$\Upsilon(11020)$}
$\Upsilon(11020)$, the candidate of $\Upsilon(6S)$, was observed in
$e^+e^-$ annihilation  in 1985~\cite{Lovelock:1985nb,Besson:1984bd}.
Its PDG mass and full width~\cite{PDG08} are
\begin{eqnarray}
M=11019\pm8\,MeV, \nonumber\\ \Gamma=79\pm16\,MeV.
\end{eqnarray}
Recently, the BaBar collaboration~\cite{:2008hx} has remeasured the
$e^+e^-\rightarrow b \bar b$ cross section by an energy scan in the
range of $10.54\,GeV$ to $11.20\,GeV$ and get the mass and full
width as
\begin{eqnarray}
M=10996\pm2\,MeV, \nonumber\\ \Gamma=37\pm3\,MeV.
\end{eqnarray}

Despite of the discrepancy in the mass and full width given by BaBar
and PDG, the observed mass is much smaller than that predicted by
the linear potential models. For example, the Cornell
model~\cite{Eichten:1978tg,Eichten:1979ms} predicted $11.14\,GeV$,
which is $121(144)\,MeV$ larger than the experimental value of
PDG(BaBar), and the modified Cornell model~\cite{Bernardini:2003ht}
gives $11.113\,GeV$, which is $104(127)\,MeV$ larger than the
experimental value of PDG(BaBar). The relativized potential model of
Godfrey and Isgur~\cite{Godfrey:1985xj} gives $11.10\,GeV$, which is
$91(114)\,MeV$ larger than the experimental value of PDG(BaBar).

Evidently, the mass of $\Upsilon(6S)$ is overestimated by the
quenched potential models by more than 100 MeV. If we take the
screening effect into account, we find, in our model, the mass of
$\Upsilon(6S)$ to be $11.023\,GeV$, which is very close to the
observed value of PDG(BaBar). The consistence of our predicted mass
with the experimental value of $\Upsilon(11020)$ indicates the
significance of the screening effect on higher excited bottomonium
states.
\subsection{Hyperfine and Fine Splittings}
We use Eq.(\ref{Hss}) to calculate the hyperfine splittings between
$\Upsilon(nS)$ and $\eta_b(nS)$, where $\tilde \delta_{\sigma}(r)$
is taken to be $\tilde \delta_{\sigma}(r) = (\sigma/\sqrt{\pi})^3\,
e^{-\sigma^2 r^2}$ as in Ref.\cite{Barnes2005}.  $\sigma$ has a
magnitude of order $m_Q$ and it represents some relativistic
smearing effects\cite{Barnes2005}.

We choose the observed splitting between $J/\psi$ and $\eta_c$ as
input to determine $\sigma$ for charmonium and have obtained a good
fit for the $\psi(3686)$-$\eta_c(3637)$ splitting~\cite{Li:2009zu}).
Here we use the observed $\eta_b$~\cite{:2008vj} and $\Upsilon(1S)$
masses to determine $\sigma$ for bottomonium, and find $\sigma$ to
be $3.3\,GeV$ in our model [see (\ref{para})]. The hyperfine
splittings for charmonium and bottomonium systems are listed in
Tab.\ref{tab:hyperfine}.  In comparison, we also list the results of
Ref.\cite{Eichten:1994gt}, which uses the Buchm\"{u}ller-Tye
potential and Ref.\cite{Godfrey:1985xj}, which uses a relativized
funnel potential.

We also list the results of the fine splittings, which are
calculated by using ~(\ref{Hls}) and ~(\ref{Ht}), between P-wave
multiplets for both charmonium and bottomonium in the same table. We
can see that our results fit the experimental values quite well and
are compatible with Ref.\cite{Eichten:1994gt} and
Ref.\cite{Godfrey:1985xj}.

\subsection{E1 transitions}
We have calculated the E1 transition widths for bottomonium using
the zeroth-order wave functions, which are marked by $SNR_0$, and
the first-order relativistically corrected wave functions, which are
marked by $SNR_1$. The results are listed in Table~\ref{E1rad},
along with the results from the potential model in
Ref.\cite{Kwong:1988ae}, in which the potential is determined by the
inverse-scattering method, for comparison.

We find our results are compatible with experimental values and
Ref.\cite{Kwong:1988ae} for most channels. Relativistic corrections
to the wave functions tend to reduce the E1 transition widths for
most channels and give better fit with experimental values. Note
that for the $\Upsilon(3S) \rightarrow\gamma\chi_{b0}$ transition
our result, $0.07(0.05)\,KeV$ with the zeroth(first)-order wave
functions, is in agreement with the experimental value
$0.061\pm0.023\,KeV$, while that of
Ref.\cite{Kwong:1988ae}($0.007\,KeV$) is too small.

But our calculated transition widths for $\Upsilon(3S)\to\gamma
\chi_{bJ} (J=1,2)$ are too large as compared with experimental data.
These may indicate that for the radially suppressed E1 transition
widths (e.g. $\Upsilon(3S)\to\gamma \chi_{bJ}(J=0,1,2)$) the
theoretical values are very sensitive to model details, and further
improvement for the model and the calculation is needed.

\subsection{Screening parameter $\mu$} We find the screening
parameter $\mu$, which represents the energy scale related to the
creation of the $Q\bar q$ and $\bar Q q$ pair or the distance when
that beyond $r\sim 1/\mu$ the screening effect becomes important, is
smaller for bottomonium~($\mu=\,0.056\,GeV$) than that for
charmonium~($\mu=\,0.0979\,GeV$)\cite{Li:2009zu}, if we try to fit
the bottomonium spectrum. We need to understand this difference of
$\mu$ between the $b\bar b$ and $c\bar c$ systems.

It is known that the string breaking is due to the creation of light
quark pairs, i.e., the formation of $Q\bar q$-$\bar Qq$ mesons. Note
that for the $b\bar b$ system the energy difference between the
$b\bar b$ bound state $\Upsilon(1S)$ and the open bottom threshold
of $B\bar B$ meson pair is 1.1 GeV, whereas for the $c\bar c$ system
the energy difference between the $c\bar c$ bound state $J/\psi$ and
the open charm threshold of $D\bar D$ meson pair is only 0.63 GeV.
This implies that for the $b\bar b$ system more energy needs to be
stored (or equivalently a longer flux tube is neded) before the
string breaking occurs than that for the $c\bar c$ system.

~~~~~~~~~~~~~~~~~~~~~~~~~~~~~~~~~~~~\\\
~~~~~~~~~~~~~~~~~~~~~~~~~~~~~~~\\

\section{Summary and Conclusions}
In this paper, as a sister work of \cite{Li:2009zu}, we incorporate
the color-screening (string breaking) effect due to light quark pair
creation into the heavy quark-antiquark long-range confinement
potential, and investigate the effects of screened potential on the
spectrum of bottomonium. We calculate the masses, electromagnetic
decays, and E1 transitions of bottomonium states in the
nonrelativistic screened potential model.

We find that the screening parameter $\mu$ is smaller for
bottomonium than that for charmonium if we try to fit the
bottomonium spectrum, and this may be understood as due to the
difference between $b\bar b$ and $c\bar c$ systems in the energy to
be stored before the string breaking occurs. The masses predicted in
the screened potential model are considerably lower for higher
bottomonium states, compared with the unscreened potential model.
Especially, the mass of $\Upsilon(6S)$ is lowered down to match that
of $\Upsilon(11020)$, whereas the linear potential model predictions
are more than 100 MeV higher than the experimental value. The fine
splittings of P-wave bottomonium states, and E1 transition rates and
leptonic decay widths are found to be compatible with experimental
data within errors.

We hope our investigation for the bottomonium system with screened
potential model will be useful in the future study of bottomonium
physics.

\section{Acknowledgement}
We would like to thank Ce Meng for many valuable discussions. This
work was supported in part by the National Natural Science
Foundation of China (No 10675003, No 10721063) and the Ministry of
Science and Technology of China (No 2009CB825200).

\begin{table*}
\caption{Experimental and theoretical mass spectrum of bottomonium
states. The experimental masses are PDG~\cite{PDG08} averages. The
masses are in units of MeV except for Ref.\cite{Godfrey:1985xj}
which is in GeV. The averaged radiuses are in units of $\rm fm$. The
results of our screened potential model are shown in comparison with
that of Ref.\cite{Godfrey:1985xj}.}
 \vspace{0.5cm}
\begin{tabular}{cc|c|cc|c}
\hline
 \multicolumn{2}{c|}{State} &Expt.&\multicolumn{2}{c|}{Theor. of ours}& Theor. of Ref~\cite{Godfrey:1985xj}\\
   &                        &     & \hspace{0.3cm}Mass\hspace{0.3cm}  & \hspace{0.3cm}$\langle r^2\rangle^{\frac{1}{2}}$\hspace{0.3cm} & \hspace{0.3cm}Mass \\
\hline
1S &  $\Upsilon(1^3{\rm S}_1) $& $ 9460.30\pm 0.26            $& 9460         &0.23       &9.46    \\
   &  $\eta_b(1^1{\rm S}_0)   $& $ 9388.9^{+3.1}_{-2.3}\pm 2.7$& 9389         &           &9.40    \\
\hline
2S &  $\Upsilon(2^3{\rm S}_1) $& $ 10023.26\pm 0.31           $& 10016        &0.52       &10.00   \\
   &  $\eta_b'(2^1{\rm S}_0)  $& $                            $& 9987         &           &9.98    \\
\hline
3S &  $\Upsilon(3^3{\rm S}_1) $& $ 10355.2\pm 0.5             $& 10351        &0.78       &10.35   \\
   &  $\eta_b(3^1{\rm S}_0)   $&                               & 10330        &           &10.34   \\
\hline
4S &  $\Upsilon(4^3{\rm S}_1) $& $ 10579.4\pm 1.2             $& 10611        &1.02       &10.63   \\
   &  $\eta_b(4^1{\rm S}_0)   $&                               & 10595        &           &        \\
\hline
5S &  $\Upsilon(5^3{\rm S}_1) $& $ 10865\pm 8                 $& 10831        &1.24       &10.88   \\
   &  $\eta_b(5^1{\rm S}_0)   $&                               & 10817        &           &        \\
\hline
6S &  $\Upsilon(6^3{\rm S}_1) $& $ 11019\pm 8                 $& 11023        &1.45       &11.10   \\
   &  $\eta_b(6^1{\rm S}_0)   $&                               & 11011        &           &        \\
\hline
7S &  $\Upsilon(7^3{\rm S}_1) $&                               & 11193        &1.66       &        \\
   &  $\eta_b(7^1{\rm S}_0)   $&                               & 11183        &           &        \\
\hline
1P &  $\chi_{b2}(1^3{\rm P}_2)  $& $ 9912.21\pm 0.26\pm 0.31  $& 9918         &0.42       &9.90    \\
   &  $\chi_{b1}(1^3{\rm P}_1 ) $& $ 9892.78 \pm 0.26\pm 0.31 $& 9897         &           &9.88    \\
   &  $\chi_{b0}(1^3{\rm P}_0)  $& $ 9859.44 \pm 0.42\pm 0.31 $& 9865         &           &9.85    \\
   &  $h_b(1^1{\rm P}_1)      $&                               & 9903         &           &9.88    \\
\hline
2P &  $\chi_{b2}(2^3{\rm P}_2)  $& $10268.65\pm0.22\pm0.50    $& 10269        &0.69       &10.26   \\
   &  $\chi_{b1}(2^3{\rm P}_1)  $& $10255.46 \pm0.22\pm0.50   $& 10251        &           &10.25   \\
   &  $\chi_{b0}(2^3{\rm P}_0)  $& $10232.5\pm0.4\pm0.5       $& 10226        &           &10.23   \\
   &  $h_c(2^1{\rm P}_1)        $&                             & 10256        &           &10.25   \\
\hline
3P &  $\chi_{b2}(3^3{\rm P}_2)  $&                             & 10540        &0.93       &        \\
   &  $\chi_{b1}(3^3{\rm P}_1)  $&                             & 10524        &           &        \\
   &  $\chi_{b0}(3^3{\rm P}_0)  $&                             & 10502        &           &        \\
   &  $h_b(3^1{\rm P}_1)        $&                             & 10529        &           &        \\
\hline
4P &  $\chi_{b2}(4^3{\rm P}_2)  $&                             & 10767        &1.15       &        \\
   &  $\chi_{b1}(4^3{\rm P}_1)  $&                             & 10753        &           &        \\
   &  $\chi_{b0}(4^3{\rm P}_0)  $&                             & 10732        &           &        \\
   &  $h_b(4^1{\rm P}_1)        $&                             & 10757        &           &        \\
\hline
5P &  $\chi_{b2}(5^3{\rm P}_2)  $&                             & 10965        &1.37       &        \\
   &  $\chi_{b1}(5^3{\rm P}_1)  $&                             & 10951        &           &        \\
   &  $\chi_{b0}(5^3{\rm P}_0)  $&                             & 10933        &           &        \\
   &  $h_b(5^1{\rm P}_1)        $&                             & 10955        &           &        \\
\hline
1D &  $\psi_3(1^3{\rm D}_3)     $&                             & 10156        &0.57       &10.16   \\
   &  $\psi_2(1^3{\rm D}_2)     $&$10161\pm0.6\pm1.6          $& 10151        &           &10.15   \\
   &  $\psi(1^3{\rm D}_1)       $& $                          $& 10145        &           &10.14   \\
   &$\eta_{c2}(1^1{\rm D}_2)    $&                             & 10152        &           &10.15   \\
\hline
2D &  $\psi_3(2^3{\rm D}_3)     $&                             & 10442        &0.82       &10.45   \\
   &  $\psi_2(2^3{\rm D}_2)     $&                             & 10438        &           &10.45   \\
   &  $\psi(2^3{\rm D}_1)       $& $                          $& 10432        &           &10.44   \\
   &$\eta_{c2}(2^1{\rm D}_2  )  $&                             & 10439        &           &10.45   \\
\hline
3D &  $\psi_3(3^3{\rm D}_3)     $&                             & 10680        &1.05       &        \\
   &  $\psi_2(3^3{\rm D}_2)     $&                             & 10676        &           &        \\
   &  $\psi(3^3{\rm D}_1)       $& $                          $& 10670        &           &        \\
   &$\eta_{c2}(3^1{\rm D}_2)    $&                             & 10677        &           &        \\
\hline
4D &  $\psi_3(4^3{\rm D}_3)     $&                             & 10886        &1.27       &        \\
   &  $\psi_2(4^3{\rm D}_2)     $&                             & 10882        &           &        \\
   &  $\psi(3^4{\rm D}_1)       $& $                          $& 10877        &           &        \\
   &$\eta_{c2}(4^1{\rm D}_2)    $&                             & 10883        &           &        \\
\hline
5D &  $\psi_3(5^3{\rm D}_3)     $&                             & 11069        &1.49       &        \\
   &  $\psi_2(5^3{\rm D}_2)     $&                             & 11065        &           &        \\
   &  $\psi(5^3{\rm D}_1)       $& $                          $& 11060        &           &        \\
   &$\eta_{c2}(5^1{\rm D}_2)    $&                             & 11066        &           &        \\
\hline \hline
\end{tabular}
\label{spectrum_table}
\end{table*}

\begin{table*}
\caption{Leptonic decay widths (in units of KeV) for bottomonium
states in our screened potential model. The widths calculated with
and without QCD corrections are marked by $\Gamma_{ee}$ and
$\Gamma^{0}_{ee}$ respectively. The experimental values are taken
from PDG~\cite{PDG08}. Predictions of two other
models\cite{Anisovich:2005jp,Pandya:2008vf} are listed for
comparison.}
 \begin{center}
 \begin{tabular}{|c|c|c|c|c|c|}
 \hline
 state            &$\Gamma^{0}_{ee}$&$\Gamma_{ee}$&Ref.\cite{Anisovich:2005jp}&Ref.\cite{Pandya:2008vf}&Exp\cite{PDG08} \\
 \hline
 $1{}^3S_{1}(9460) $      &2.31     &1.60          &1.314                     &1.320       &$1.340\pm0.018$\\
 \hline
 $2{}^3S_{1}(10023)$      &0.92     &0.64          &0.576                     &0.628       &$0.612\pm0.011$\\
 \hline
 $3{}^3S_{1}(10355)$      &0.64     &0.44          &0.476                     &0.263       &$0.443\pm0.008$\\
 \hline
 $4{}^3S_{1}(10579)$      &0.51     &0.35          &0.248                     &0.104       &$0.272\pm0.029$ \\
 \hline
 $5{}^3S_{1}(10865)$      &0.42     &0.29          &0.310                     &0.04        &$0.31\pm0.07$\\
\hline
 $6{}^3S_{1}(11019)$      &0.37     &0.25          &                          &            &$0.130\pm0.030$ \\
 \hline
 $7{}^3S_{1}(11193)$      &0.32     &0.22          &                          &            & \\
 \hline
\end{tabular}
\end{center}
\label{tab:ee}
\end{table*}

\begin{table*}
\caption{Two-photon decay widths (in units of eV) of the
pseudoscalar
  (${}^1\!S_0$), scalar (${}^3\!P_0$), and tensor (${}^3\!P_2$) bottomonium states. Bottomonium masses are in units of
  MeV.}
\begin{ruledtabular}
\begin{tabular}{c|c|cccccccc |c  }
      &                    & \multicolumn{8}{c|}{Theory}                                                                                  & Experiment\\
\hline
 state&  mass              &Ref.\cite{Ebert:2003mu} &Ref.\cite{Munz:1996hb} &Ref.\cite{Godfrey:1985xj}
                           &Ref.\cite{Bhaduri:1981pn}&Ref.\cite{Resag:1994ki} &Ref.\cite{Gupta:1996ak}
                           &Ref.\cite{Schuler:1997yw} &Ours   &PDG~\cite{PDG08} \\
\hline
 $\eta_b(1{}^1\!S_0)$&9389 &350        &220     &214      &266     &192    &460  &460   &527   &\\
\hline
 $\eta_b(2{}^1\!S_0)$&9987 &150        &110     &121      &95.0    &116    &     &200   &263   & \\
\hline
$\eta_b(3{}^1\!S_0)$&10330 &100        &84      &90.6     &67.9    &93.5   &     &      &172  & \\
\hline
$\eta_b(4{}^1\!S_0)$&10595 &           &71      &75.5     &56.3    &81.8   &     &      &105   &\\
\hline
$\eta_b(5{}^1\!S_0)$&10817 &           &        &         &        &       &     &      &121   & \\
\hline
$\eta_b(6{}^1\!S_0)$&11011 &           &        &         &        &       &     &      &50  & \\
\hline
 $\chi_{b0}(1{}^3\!P_0)$&9859&38       &24      &20.8     &27.3    &24.1   &80   &43    &37  & \\
\hline
$\chi_{b0}(2{}^3\!P_0)$&10233&29       &26      &22.7     &26.9    &27.3   &     &      &37 & \\
\hline
$\chi_{b0}(3{}^3\!P_0)$&10502&         &        &         &        &       &     &      &35  & \\
\hline
 $\chi_{b2}(1{}^3\!P_2)$&9912&8        &5.6     &5.14     &2.56    &6.45   &8    &7.4   &6.6  &\\
\hline
$\chi_{b2}(2{}^3\!P_2)$&10269&6        &6.8     &6.21     &6.11    &8.1    &     &      &6.7  & \\
\hline
$\chi_{b2}(3{}^3\!P_2)$&10540&         &        &         &        &       &     &      &6.4  & \\
\end{tabular}
\end{ruledtabular}
\label{twophoton}
\end{table*}

\begin{table*}
\caption{E1 transition rates of bottomonium states in our screened
potential model~(those calculated by the zeroth-order wave functions
are marked by $SNR_0$ and those by the first-order relativistically
corrected wave functions are marked by $SNR_1$). We also list the
results of one potential model, in which the potential is determined
by the inverse-scattering method, for
comparison~\cite{Kwong:1988ae}.} \vskip 0.3cm
\begin{ruledtabular}
\begin{tabular}{l| ll| c c| c c c| c c  }
state & Initial meson  & Final meson
&\multicolumn{2}{c}{E$_{\gamma}$ (MeV)} &
\multicolumn{3}{c}{$\Gamma_{\rm thy}$~(keV)}
& $\Gamma_{\rm expt}$~(keV) & \\
            &                                &                            &Ref~\cite{Kwong:1988ae}&$SNR_{0(1)}$&Ref~\cite{Kwong:1988ae}&$SNR_0$ &$SNR_1$ &PDG~\cite{PDG08} \\
\hline
2S $\to$ 1P &$\Upsilon(2^3{\rm S}_1)(10023) $&$\chi_{b2}(1^3{\rm P}_2)   $&110   &110         &2.14 &2.62    &2.46      &$2.29\pm0.23$\\
            &                                &$\chi_{b1}(1^3{\rm P}_1)   $&131   &130         &2.18 &2.54    &2.08      &$2.21\pm0.22$ \\
            &                                &$\chi_{b0}(1^3{\rm P}_0)   $&162   &163         &1.39 &1.67    &1.11      &$1.22\pm0.16$ \\
            &$\eta_c(2^1{\rm S}_0)(9389)    $&$h_b(1^1{\rm P}_1)         $&      &83          &     &6.10    &5.57      &\\
\hline
3S $\to$ 2P &$\Upsilon(3^3{\rm S}_1)(10355) $&$\chi_{b2}(2^3{\rm P}_2)   $&87    &86          &2.78 &3.23    &3.04      &$2.66\pm0.41$   \\
            &                                &$\chi_{b1}(2^3{\rm P}_1)   $&99    &99          &2.52 &2.96    &2.44      &$2.56\pm0.34$    \\
            &                                &$\chi_{b0}(2^3{\rm P}_0)   $&124   &122         &1.65 &1.83    &1.23      &$1.20\pm0.16$    \\
            &$\eta_c(3^1{\rm S}_0)(10330)   $&$h_b(2^1{\rm P}_1)         $&      &74          &     &11.0    &10.1      &\\
\hline

3S $\to$ 1P &$\Upsilon(3^3{\rm S}_1)        $&$\chi_{b2}(1^3{\rm P}_2)   $&433   &434         &0.025&0.25    &1.26      &$<0.386\pm0.035$   \\
            &                                &$\chi_{b1}(1^3{\rm P}_1)   $&453   &452         &0.017&0.17    &0.14      &$<0.0345\pm0.0031$  \\
            &                                &$\chi_{b0}(1^3{\rm P}_0)   $&484   &484         &0.007&0.07    &0.05      &$0.061\pm0.023$    \\
            &$\eta_c(3^1{\rm S}_0)          $&$h_b(1^1{\rm P}_1)         $&      &418         &     &1.24    &5.68    &\\
\hline
4S $\to$ 3P &$\Upsilon(4^3{\rm S}_1)(10579) $&$\chi_{b2}(3^3{\rm P}_2)   $&      &40          &     &0.55    &0.52      &   \\
            &                                &$\chi_{b1}(3^3{\rm P}_1)   $&      &55          &     &0.91    &0.74      &    \\
            &                                &$\chi_{b0}(3^3{\rm P}_0)   $&      &77          &     &0.82    &0.54      &    \\
            &$\eta_c(4^1{\rm S}_0)(10595)   $&$h_b(3^1{\rm P}_1)         $&      &67          &     &14.3    &12.9     &\\
\hline
4S $\to$ 2P &$\Upsilon(4^3{\rm S}_1)        $&$\chi_{b2}(2^3{\rm P}_2)   $&      &306         &     &0.14    &0.56      &   \\
            &                                &$\chi_{b1}(2^3{\rm P}_1)   $&      &319         &     &0.09    &0.001     &    \\
            &                                &$\chi_{b0}(2^3{\rm P}_0)   $&      &341         &     &0.04    &0.21      &    \\
            &$\eta_c(4^1{\rm S}_0)          $&$h_b(2^1{\rm P}_1)         $&      &334         &     &0.95    &2.16    &\\
\hline
4S $\to$ 1P &$\Upsilon(4^3{\rm S}_1)        $&$\chi_{b2}(1^3{\rm P}_2)   $&      &646         &     &0.15    &0.86      &   \\
            &                                &$\chi_{b1}(1^3{\rm P}_1)   $&      &664         &     &0.10    &0.20     &    \\
            &                                &$\chi_{b0}(1^3{\rm P}_0)   $&      &695         &     &0.04    &0.001      &    \\
            &$\eta_c(4^1{\rm S}_0)          $&$h_b(1^1{\rm P}_1)         $&      &669         &     &0.90    &5.64     &\\
\hline
1P $\to$ 1S &$\chi_{b2}(1^3{\rm P}_2)(9912) $&$\Upsilon(1^3{\rm S}_1)(9460)$&443 &442         &37.8 &38.2    &32.6     &    \\
            &$\chi_{b1}(1^3{\rm P}_1)(9893) $&                            &443   &423         &32.8 &33.6    &30.0     &    \\
            &$\chi_{b0}(1^3{\rm P}_0)(9859) $&                            &392   &391         &26.1 &26.6    &24.3     &    \\
            &$h_b(1^1{\rm P}_1)(9903)       $&$\eta_b(1^1{\rm S}_0)(9389)$&      &501         &     &55.8    &36.3     &  \\
\hline
2P $\to$ 2S &$\chi_{b2}(2^3{\rm P}_2)(10269)$&$\Upsilon(2^3{\rm S}_1)(10023)$&242&243         &18.7 &18.8    &14.2     &    \\
            &$\chi_{b1}(2^3{\rm P}_1)(10255)$&                            &230   &230         &15.9 &15.9    &13.8     &    \\
            &$\chi_{b0}(2^3{\rm P}_0)(10233)$&                            &205   &207         &11.3 &11.7    &11.6     &    \\
            &$h_b(2^1{\rm P}_1)(10256)      $&$\eta_b(2^1{\rm S}_0)(9987)$&      &266         &     &24.7    &15.3     &  \\
\hline
2P $\to$ 1S &$\chi_{b2}(2^3{\rm P}_2) $&$\Upsilon(1^3{\rm S}_1)          $&777   &777         &9.75 &13.0    &12.5     &   \\
            & $\chi_{b1}(2^3{\rm P}_1)$&                                  &765   &764         &9.31 &12.4    &8.56     &   \\
            & $\chi_{b0}(2^3{\rm P}_0)$&                                  &742   &743         &8.48 &11.4    &4.50     &    \\
            & $h_b(2^1{\rm P}_1)$      &$\eta_b(1^1{\rm S}_0)$            &      &831         &     &15.9    &18.0     &  \\
\hline
3P $\to$ 3S &$\chi_{b2}(3^3{\rm P}_2)(10540)$&$\Upsilon(3^3{\rm S}_1)    $&170   &183         &12.1 &15.6    &11.1     &   \\
            &$\chi_{b1}(3^3{\rm P}_1)(10524)$&                            &159   &167         &10.1 &12.0    &9.97     &    \\
            &$\chi_{b0}(3^3{\rm P}_0)(10502)$&                            &144   &146         &7.46 &7.88    &7.67     &    \\
            &$h_b(3^1{\rm P}_1)(10529)      $&$\eta_b(3^1{\rm S}_0)      $&      &196         &     &19.2    &11.6     &  \\
\hline
3P $\to$ 2S &$\chi_{b2}(3^3{\rm P}_2)(10540)$&$\Upsilon(2^3{\rm S}_1)    $&491   &504         &3.78 &6.00    &6.89     &    \\
            &$\chi_{b1}(3^3{\rm P}_1)(10524)$&                            &481   &489         &3.56 &5.48    &5.39     &    \\
            &$\chi_{b0}(3^3{\rm P}_0)(10502)$&                            &466   &468         &3.24 &4.80    &3.67     &    \\
            &$h_b(3^1{\rm P}_1)(10529)      $&$\eta_b(2^1{\rm S}_0)      $&      &528         &     &6.89    &10.3     &  \\
\hline
3P $\to$ 1S &$\chi_{b2}(3^3{\rm P}_2)(10540)$&$\Upsilon(1^3{\rm S}_1)    $&1012  &1024        &3.80 &7.09    &6.76     &    \\
            &$\chi_{b1}(3^3{\rm P}_1)(10524)$&                            &1003  &1010        &3.69 &6.80    &3.39     &    \\
            &$\chi_{b0}(3^3{\rm P}_0)(10502)$&                            &989   &990         &3.54 &6.41    &0.86     &   \\
            &$h_b(3^1{\rm P}_1)(10529)      $&$\eta_b(1^1{\rm S}_0)      $&      &1078        &     &8.27    &9.46     &  \\
\hline
2P $\to$ 1D &$\chi_{b2}(2^3{\rm P}_2)    $&$\Upsilon(1^3{\rm D}_3)(10156)$&107   &113         &2.62 &3.33    &3.13       &\\
            &                             &$\Upsilon(1^3{\rm D}_2)(10151)$&112   &117         &0.54 &0.66    &0.58        &\\
            &                             &$\Upsilon(1^3{\rm D}_1)(10145)$&119   &123         &0.043&0.05    &0.04       &\\
            &$\chi_{b1}(2^3{\rm P}_1)    $&$\Upsilon(1^3{\rm D}_2)       $&99    &104         &1.86 &2.31    &2.26       &  \\
            &                             &$\Upsilon(1^3{\rm D}_1)       $&106   &110         &0.76 &0.92    &0.84       & \\
            &$\chi_{b0}(2^3{\rm P}_0)    $&$\Upsilon(1^3{\rm D}_1)       $&81    &87          &1.36 &1.83    &1.85       &  \\
            &$h_b(2^1{\rm P}_1)          $&$h_{b2}(1^1{\rm D}_2)(10152)  $&      &104         &     &7.74    &7.42       &\\
\hline
3P $\to$ 2D &$\chi_{b2}(3^3{\rm P}_2)    $&$\Upsilon(2^3{\rm D}_3)(10442)$&82   &97           &3.01 &5.05    &4.69       &\\
            &                             &$\Upsilon(2^3{\rm D}_2)(10438)$&85   &101          &0.61 &1.02    &0.89        &\\
            &                             &$\Upsilon(2^3{\rm D}_1)(10432)$&91   &107          &0.05 &0.08    &0.07       &\\
            &$\chi_{b1}(3^3{\rm P}_1)    $&$\Upsilon(2^3{\rm D}_2)       $&75   &86           &2.08 &3.10    &2.98       &  \\
            &                             &$\Upsilon(2^3{\rm D}_1)       $&81   &92           &0.86 &1.26    &1.13       & \\
            &$\chi_{b0}(3^3{\rm P}_0)    $&$\Upsilon(2^3{\rm D}_1)       $&66   &70           &1.85 &2.23    &2.21       &  \\
            &$h_b(3^1{\rm P}_1)          $&$h_{b2}(2^1{\rm D}_2)(10439)  $&     &89           &     &11.8    &11.2       &\\
\hline
\end{tabular}
\end{ruledtabular}
\label{E1rad}
\end{table*}

\begin{table*}
\vskip 0.3cm
\begin{ruledtabular}
\begin{tabular}{l| ll| c c| c c c| c c  }
state & Initial meson  & Final meson
&\multicolumn{2}{c}{E$_{\gamma}$ (MeV)} &
\multicolumn{3}{c}{$\Gamma_{\rm thy}$~(keV)}
& $\Gamma_{\rm expt}$~(keV) & \\
            &                                &                            &Ref~\cite{Kwong:1988ae}&$SNR_{0(1)}$&Ref~\cite{Kwong:1988ae}&$SNR_0$ &$SNR_1$ &PDG~\cite{PDG08} \\
\hline
3P $\to$ 1D &$\chi_{b2}(3^3{\rm P}_2)    $&$\Upsilon(1^3{\rm D}_3)(10156)$&      &377         &$\approx 0$  &$\approx 0$ &0.05       &\\
            &                             &$\Upsilon(1^3{\rm D}_2)(10151)$&      &381         &$\approx 0$  &$\approx 0$ &$\approx 0$        &\\
            &                             &$\Upsilon(1^3{\rm D}_1)(10145)$&      &387         &$\approx 0$  &$\approx 0$ &$\approx 0$       &\\
            &$\chi_{b1}(3^3{\rm P}_1)    $&$\Upsilon(1^3{\rm D}_2)       $&      &366         &$\approx 0$  &$\approx 0$ &0.09       &  \\
            &                             &$\Upsilon(1^3{\rm D}_1)       $&      &372         &$\approx 0$  &$\approx 0$ &0.004       & \\
            &$\chi_{b0}(3^3{\rm P}_0)    $&$\Upsilon(1^3{\rm D}_1)       $&      &351         &$\approx 0$  &$\approx 0$ &0.17       &  \\
            &$h_b(3^1{\rm P}_1)          $&$h_{b2}(1^1{\rm D}_2)(10152)  $&      &370         &             &$\approx 0$ &0.24       &\\
\hline
1D $\to$ 1P &$\Upsilon(1^3{\rm D}_3)(10156)$&$\chi_{b2}(1^3{\rm P}_2)    $&245   &240         &24.3 &26.4    &24.5       &\\
            &$\Upsilon(1^3{\rm D}_2)(10151)$&$\chi_{b2}(1^3{\rm P}_2)    $&240   &236         &5.7  &6.29    &5.87        &\\
            &                             &$\chi_{b1}(1^3{\rm P}_1)      $&261   &255         &22.0 &23.8    &19.8       &\\
            &$\Upsilon{\rm D}_1)(10145)  $&$\chi_{b2}(1^3{\rm P}_2)      $&233   &230         &0.58 &0.65    &0.61       &  \\
            &                             &$\chi_{b1}(1^3{\rm P}_1)      $&254   &249         &11.3 &12.3    &10.3       & \\
            &                             &$\chi_{b0}(1^3{\rm P}_0)      $&285   &282         &21.4 &23.6    &16.7       &  \\
            &$h_{b2}(1^1{\rm D}_2)(10152)$&$h_b(1^1{\rm P}_1)            $&      &246         &     &42.3    &36.5       &\\
\hline
2D $\to$ 2P &$\Upsilon(2^3{\rm D}_3)(10442)$&$\chi_{b2}(2^3{\rm P}_2)    $&174   &172         &16.3 &18.0    &15.9       &\\
            &$\Upsilon(2^3{\rm D}_2)(10438)$&$\chi_{b2}(2^3{\rm P}_2)    $&171   &168         &3.83 &4.17    &3.82       &\\
            &                             &$\chi_{b1}(2^3{\rm P}_1)      $&183   &181         &14.2 &15.7    &12.1       &\\
            &$\Upsilon{\rm D}_1)(10432)  $&$\chi_{b2}(2^3{\rm P}_2)      $&165   &162         &0.38 &0.42    &0.39       &  \\
            &                             &$\chi_{b1}(2^3{\rm P}_1)      $&178   &175         &7.2  &7.87    &6.35        & \\
            &                             &$\chi_{b0}(2^3{\rm P}_0)      $&202   &198         &14.2 &15.1    &9.49      &  \\
            &$h_{b2}(2^1{\rm D}_2)(10439)$&$h_b(2^1{\rm P}_1)            $&      &181         &     &31.3    &25.4       &\\
\hline
2D $\to$ 1P &$\Upsilon(1^3{\rm D}_3)     $&$\chi_{b2}(1^3{\rm P}_2)      $&518   &517         &3.94 &4.01    &3.73       &\\
            &$\Upsilon(1^3{\rm D}_2)     $&$\chi_{b2}(1^3{\rm P}_2)      $&514   &513         &0.97 &0.98    &0.68        &\\
            &                             &$\chi_{b1}(1^3{\rm P}_1)      $&534   &531         &3.25 &3.26    &4.46       &\\
            &$\Upsilon{\rm D}_1)         $&$\chi_{b2}(1^3{\rm P}_2)      $&509   &507         &0.10 &0.11    &0.05       &  \\
            &                             &$\chi_{b1}(1^3{\rm P}_1)      $&529   &525         &1.75 &1.76    &1.87        & \\
            &                             &$\chi_{b0}(1^3{\rm P}_0)      $&559   &557         &2.76 &2.79    &6.20      &  \\
            &$h_{b2}(1^1{\rm D}_2)       $&$h_b(1^1{\rm P}_1)            $&      &522         &     &6.19    &7.30      &\\
\end{tabular}
\end{ruledtabular}
\label{E1rad}
\end{table*}

\begin{table*}
\caption{Hyperfine and fine splittings in units of MeV for
charmonium and bottomonium in our model. Here $\sigma$ is
$1.362\,GeV$ for charmonium and $3.3\,GeV$ for bottomonium. The
experimental values are the mass differences of the corresponding
charmonium and bottomonium states taken from PDG~\cite{PDG08}.
 Results of Ref.\cite{Eichten:1994gt} and
Ref.\cite{Godfrey:1985xj} are listed for comparison.}
 \begin{center}
\begin{tabular}{|@{\hspace{0.5cm}}c@{\hspace{0.5cm}}|@{\hspace{0.5cm}}c@{\hspace{0.5cm}}c@{\hspace{0.5cm}}|@{\hspace{0.5cm}}c@{\hspace{0.5cm}}c@{\hspace{0.5cm}}c@{\hspace{0.5cm}}c|}
 \hline
 State              & \multicolumn{2}{c|}{Charmonium} &\multicolumn{4}{c|}{Bottomonium}\\
                    & Ours            & Exp          & Ours  &Ref.\cite{Eichten:1994gt}&Ref.\cite{Godfrey:1985xj} & Exp          \\
 \hline
 $1^3S_1-1^1S_0$    &118              &$116.6\pm1.2$ &71     &87     &60           &$71.4^{+2.3}_{-3.1}\pm3.7$   \\
 \hline
 $2^3S_1-2^1S_0$    &50               &$52\pm4$      &29     &44     &20&\\
\hline
 $3^3S_1-3^1S_0$    &31               &              &21     &41     &10&\\
\hline
 $1^3P_2-1^3P_1$    &44               &$45.54\pm0.11$&21     &22     &20&$19.43\pm0.57$   \\
\hline
 $1^3P_1-1^3P_0$    &77               &$95.91\pm0.32$&32     &30     &30&$33.34\pm0.66$   \\
\hline
 $2^3P_2-2^3P_1$    &36               &              &18     &18     &10&$13.19\pm0.77$   \\
\hline
 $2^3P_1-2^3P_0$    &59               &              &25     &25     &20&$22.96\pm0.84$   \\
\hline
 $3^3P_2-3^3P_1$    &30               &              &16     &       &&   \\
\hline
 $3^3P_1-3^3P_0$    &47               &              &22     &       &&   \\
 \hline
\end{tabular}
\end{center}
\label{tab:hyperfine}
\end{table*}

\end{document}